\documentclass[12pt]{article}
\usepackage{graphicx,epsf,subfigure}
\usepackage{pstricks,pst-node,psfrag}
\usepackage{amsthm,amssymb,amsmath}
\begin{document}

\begin{center}
Working Paper

\large \textbf{Propensity Score Analysis with Matching Weights} \vspace{1cm} 

\normalsize
Liang Li, Ph.D. 

Associate Staff of Biostatistics \\
Department of Quantitative Health Sciences, Cleveland Clinic \\
9500 Euclid Avenue, Cleveland, Ohio, 44195, U.S.A. \\
E-mail: lil2@ccf.org

May 13, 2011 
\end{center}

\begin{abstract}
The propensity score analysis is one of the most widely used methods for studying the causal treatment effect in observational studies. This paper studies treatment effect estimation with the method of matching weights. This method resembles propensity score matching but offers a number of new features including efficient estimation, rigorous variance calculation, simple asymptotics, statistical tests of balance, clearly identified target population with optimal sampling property, and no need for choosing matching algorithm and caliper size. In addition, we propose the mirror histogram as a useful tool for graphically displaying balance. The method also shares some features of the inverse probability weighting methods, but the computation remains stable when the propensity scores approach 0 or 1. An augmented version of the matching weight estimator is developed that has the double robust property, i.e., the estimator is consistent if either the outcome model or the propensity score model is correct. In the numerical studies, the proposed methods demonstrated better performance than many widely used propensity score analysis methods such as stratification by quintiles, matching with propensity scores, and inverse probability weighting.
\end{abstract}


\footnotetext[1]{This paper will be presented at the 2011 Atlantic Causal Inference Conference, Ann Arbor, Michigan}

\newpage

\section{Introduction}
\label{sec:intro}
Propensity score analysis is an important statistical tool for adjusting for confounding in observational studies \cite{RosenbaumRubin1983}, and has been widely used across many research fields such as epidemiology, economics, political and social sciences \cite{DAgostino1998, Imbens2004, Ho2007}. In this paper, we study the population propensity score analysis \cite{RobinThomas1996}. Let $\{ Y_i, Z_i, \mathbf{X}_i, i=1,2,...n \}$ be the observed data from $n$ independent subjects randomly sampled from a population of research interest, where $Y_i$ denotes the outcome of research interest, $Z_i = 1$ or $0$ indicates whether the subject was assigned to the treatment or control, and $\mathbf{X}_i$ is a vector of variables related to the treatment assignment and the outcome. The research question is to study whether the treatment has an effect on the outcome and to estimate that effect quantitatively.

It is helpful to conceptualize this problem using the potential outcomes framework. We assume that the observed outcome for subject $i$ would be $Y_{1i}$ if the subject had been assigned to the treatment group and $Y_{0i}$ if the control group. Since the subject can only receive either the treatment or the control, $Y_{1i}$ and $Y_{0i}$ are potential outcomes that are never observed simultaneously in reality. Their relationship with the observed outcome $Y_i$ and assignment $Z_i$ is assumed to be: $Y_i = Y_{1i}Z_i + Y_{0i}(1-Z_i)$. This relationship is called the Stable Unit Treatment Value Assumption \cite{SUTVA}. It implies that the observed outcome of a subject is solely determined by the potential outcomes and the assignment for that subject, and does not interfere with data from other subjects. This assumption is satisfied in the setting considered in this paper as we assume that the subjects are independent. Another assumption needed for propensity score analysis is the assumption of ``no unmeasured confounders'' or ``strongly ignorable treatment assignment'': $ ( Y_{1i}, Y_{0i} ) \perp Z_i ~ \mid ~ \mathbf{X}_i  $. It states that $\mathbf{X}_i$ must include all relevant variables such that conditional on these observed variables, the potential outcomes are independent of the treatment assignment. Notation $\perp$ denotes independence. Variables in $\mathbf{X}_i$ are called confounders and this assumption requires that there should be no unmeasured confounders.

The goal of the propensity score analysis is to estimate the effect of the treatment, which may be defined for each subject as $\Delta_i = E( Y_{1i} - Y_{0i} )$, the difference in expected potential outcomes of the same subject. If $\Delta_i \equiv \Delta$, a typical setting studied in many numerical studies \cite{Lunceford2004, Austin2009}, the treatment effect is homogeneous. If $\Delta_i$ may be different for different subjects, the treatment effect is heterogeneous, and one may study the average causal effect $\Delta_0 = E(\Delta_i)$, where the expectation is taken over some target population of research interest. We consider both cases in this paper and when the treatment effect is heterogeneous, we assume $\Delta_i = E( Y_{1i} - Y_{0i} \mid \mathbf{X}_i ) = \Delta(\mathbf{X}_i)$, for a function $\Delta(.)$ of the confounders.

The propensity score is defined for each subject to be the conditional probability of receiving the treatment, given confounders, i.e., $ e_i = \mathrm{Pr}( Z_i = 1 \mid \mathbf{X}_i ) $. Rosenbaum and Rubin \cite{RosenbaumRubin1983} proved $ \mathbf{X}_i \perp Z_i ~\mid~ e_i $ and $ ( Y_{1i}, Y_{0i} ) \perp Z_i ~\mid~ e_i $, which imply that the treatment may be viewed as being randomly assigned to subjects with the same propensity score. Therefore, one can intuitively think of the entire data set as a collection of many tiny randomized experiments, each defined on a distinct value of the propensity score. An estimator for the causal treatment effect may be formed by properly aggregating results from these tiny experiments.

Popular methods that use propensity score to estimate the treatment effect include stratification or regression \cite{RosenbaumRubin1984, DAgostino1998}, matching \cite{RosenbaumRubin1985, Austin2008}, inverse probability weighting \cite{Lunceford2004}, or a combination of them \cite{Ho2007}. This paper studies a new approach, the method of matching weights. Section \ref{sec:MW} introduces the estimator, and discusses its asymptotic properties, estimand, computation, and balance diagnosis. It shares some features of both the propensity score matching and the inverse probability weighting and avoids some of their drawbacks. Section \ref{sec:DR} developed an augmented matching weight estimator that is double robust and efficient within a class of asymptotically linear estimators. The theoretical development parallels that of the inverse probability weighting method \cite{Robins1994}. Section \ref{sec:simu} presents numerical studies to compare the proposed estimators with competing estimators.

\section{Matching Weight Estimator}
\label{sec:MW}
The propensity score $e_i$ is often estimated by a logistic regression of $Z_i$ on $\mathbf{X}_i$:
\begin{equation}
e_i = e(\mathbf{X}_i, \beta) = \mathrm{Pr}( Z_i = 1 \mid \mathbf{X}_i ) = \frac{ \exp\{ \mathbf{X}_i^T \beta \} }{ 1 + \exp\{ \mathbf{X}_i^T \beta \} }. \label{eq:PS}
\end{equation}
We call (\ref{eq:PS}) the propensity score model. Throughout this paper, the term ``propensity score'' refers to $e_i$ on its probability scale, i.e., $0 < e_i < 1$, unless otherwise specified. The propensity score can not be $0$ or $1$, otherwise the subject can not potentially be assigned to both treatments, and one of the potential outcomes is undefined.

We define the matching weight for subject $i$ as
\begin{equation}
W_i = \frac{ \mathrm{min}( 1-e_i, e_i ) }{ Z_i e_i + (1-Z_i)(1-e_i) }. \label{eq:MW}
\end{equation}
The matching weight estimator is
\begin{equation}
\hat{\Delta}_{MW} = \frac{ \sum_{i=1}^n W_i Z_i Y_i }{ \sum_{i=1}^n W_i Z_i } - \frac{ \sum_{i=1}^n W_i (1-Z_i) Y_i }{ \sum_{i=1}^n W_i (1-Z_i) }. \label{eq:MWestimator}
\end{equation}
The matching weight is a modification of the inverse probability weight with $\mathrm{min}(1-e_i, e_i)$ placed in the numerator, which prevents the weight to be excessively large when $e_i$ approaches $0$ or $1$, and stabilizes the estimator and improves its efficiency. Here is the intuition behind this estimator. Suppose we focus on a small stratum of $m_0$ subjects with propensity scores very close to $e_0$. Then we would expect that there are roughly $m_0 e_0$ treated subjects and $m_0 (1-e_0)$ controls. When $e_0 \leq 0.5$, there are more controls than the treated subjects in this stratum, hence we give less weights to the controls in order to achieve balance. When $e_0 > 0.5$, there are more treated subjects than controls, and we give less weights to the treated subjects.

Another modification is to give the treated subjects weight $1$, and controls weight $e_i/(1-e_i)$, leading to an estimator of the average treatment for the treated \cite{Hirano2001}. However, this weight may still be very large and unstable when $e_i$ is close to $1$, reflecting the difficulty to recover information about $Y_{0i}$ when mostly likely we can only observe $Y_{1i}$.

Since the matching weight is between $0$ and $1$, it can be viewed as a sampling probability: the treated and control subjects are sampled with sampling probabilities that depend on the propensity scores. Consequently, $\hat{\Delta}_{MW}$ is the average difference in mean outcomes of the sampled subjects. We define the effective sample size of the sampled subjects in the treatment group as $\sum_{i=1}^n W_i Z_i$ and, for the controls, $\sum_{i=1}^n W_i (1-Z_i)$. They are asymptotically equal. The following result further characterizes the property of the populations sampled by the matching weights.

\textbf{Proposition 1.} \textit{ Let $f(e)$ be the density function of the propensity score $e$, and $S_0(e)$ and $S_1(e)$ be the sampling probabilities for the subjects with $Z=0$ and $Z=1$ such that both the expected effective sample sizes and the distributions of the propensity score are asymptotically equal between the sampled subgroups with $Z=0$ and $Z=1$. Then
$$
S_0(e) \leq \mathrm{min}(1-e, e)/(1-e) ~~\mathrm{and}~~ S_1(e) \leq \mathrm{min}(1-e, e)/e
$$
and the equality holds simultaneously.} 

This result shows that the matching weight is optimal in the sense that it maximizes the sizes of the sampled subgroups while keeping them balanced in their effective sample sizes and their distributions of the propensity score, and hence $\mathbf{X}$. The density function of the propensity score is identical for the two subpopulations sampled by the matching weights:
$$
f^*(e) = \frac{ f(e)\mathrm{min}(1-e, e) }{ \int f(u)\mathrm{min}(1-u, u) du}, ~~~~ 0 < e < 1.
$$
We call them maximal balanced subpopulations. The matching weight is very similar to propensity score matching. First, they both produce weighted subgroups that have similar distributions in the propensity score and confounders, and similar effective sample sizes. Second, they both let each subject in the data to be under-represented in the sense that they are weighted by a number that is non-negative and no more than $1$. One difference is that with matching, each subject receives a weight of $1$ (matched) or $0$ (unmatched), while with matching weight, all subjects are retained and we deal with their probabilities of being matched instead of deciding who can be matched and who can not. Another difference is that matching weight is calculated for each subject independently, but matching derives the weights from a matching algorithm, which may introduce complicated dependency between matched subjects. Alternative weighting methods, such as the inverse probability weighing, allow each subject in the data to be over-represented in the sense that their weights are bigger than $1$.

\textbf{Proposition 2.} \textit{ Assume that the propensity score model is known. When $n \rightarrow \infty$, we have
$$
\hat{\Delta}_{MW} \rightarrow_p \frac{ E\{ \mathrm{min}(1-e_i, e_i)\Delta(\mathbf{X}_i) \} }{ E\{ \mathrm{min}(1-e_i, e_i) \} } \equiv \Delta_0
~~\mathrm{and}~~ \sqrt{n}( \hat{\Delta}_{MW} - \Delta_0 ) \rightarrow_d N(0, V_{MW})
$$
where
$$
V_{MW} = \frac{ E\Bigl\{ \bigl[ \mathrm{min}(1-e_i, e_i) ( Y_{1i} - \mu_1 ) \bigr]^2/e_i + \bigl[ \mathrm{min}(1-e_i, e_i) ( Y_{0i} - \mu_0 ) \bigr]^2/(1-e_i) \Bigr\} }{ \Bigl\{ E[ \mathrm{min}(1-e_i, e_i) ] \Bigr\}^2 }
$$
with $
\mu_1 = \int E( Y_{1i} \mid e_i )f^*(e_i) de_i ~~\mathrm{and}~~ \mu_0 = \int E( Y_{0i} \mid e_i )f^*(e_i) de_i.
$ } 

This result establishes the asymptotic distribution of the matching weight estimator. Under the special case $\Delta(\mathbf{X}) \equiv \Delta$, $\Delta_0 = \Delta$ and the matching weight estimator consistently estimates the treatment effect. Under heterogeneous conditions, its estimand is the average treatment effect over the maximal balanced subpopulations. In terms of the estimand, the matching weight method is again similar to propensity score matching, but with an advantage: the subpopulation on which the average treatment effect is defined is unique and optimal in the sense of Proposition 1. In propensity score matching, often a subset of the treated and control subjects are selected into the matched data, and the distribution of matched data depends on the matching algorithm and the caliper size \cite{AustinCaliper}. It is unclear what are the subpopulations under comparison. The inverse probability weighting has the marginal structural model interpretation \cite{Robins2000} and it estimates the average causal effect over the entire population under both homogeneous and heterogeneous conditions. However, this goal is achieved at some cost. Although it is undesirable to include in the study subjects that can almost only be assigned to the treatment or control, in practical situations, the propensity score is often unknown and must be calculated from a parsimonious mathematical model. Sometimes the calculated propensity scores are very close to $0$ or $1$. In such case, the method has to use few treated or control subjects to ``recover'' the virtual populations in which all subjects received the control, or the treatment. This is done by weighting the data with the inverse of small probabilities. Data sparsity like this may result in large loss of efficiency and unstable calculation \cite{Kang}. Under homogeneous conditions, there is no need to estimate the marginal mean of the potential outcomes in order to get to an estimator of the treatment effect. In this case, the estimands of the matching weight and inverse probability weighting methods are identical.

We can estimate the matching weights jointly with the treatment effect by solving the following estimating equations with respect to $\mathbf{\theta} = ( \mu_1, \mu_0, \mathbf{\beta}^T )^T$:
\begin{equation}
\mathbf{0} = \sum_{i=1}^n \mathbf{\phi}_i( \mathbf{\theta} ) = \sum_{i=1}^n \left[
\begin{array}{c}
W( \mathbf{X}_i, Z_i, \mathbf{\beta} ) Z_i ( Y_i - \mu_1 ) \\
W( \mathbf{X}_i, Z_i, \mathbf{\beta} ) (1-Z_i)(Y_i - \mu_0) \\
\frac{ Z_i - e(\mathbf{X}_i, \mathbf{\beta}) }{ e(\mathbf{X}_i, \mathbf{\beta})\bigl[ 1 - e(\mathbf{X}_i, \mathbf{\beta}) \bigr] } e_{\mathbf{\beta}}(\mathbf{X}_i)
\end{array} \right] \label{eq:MW.EE}
\end{equation}
where $e_{\mathbf{\beta}}(\mathbf{X}_i) = \partial e(\mathbf{X}_i, \mathbf{\beta})/\partial \mathbf{\beta} $ and we rewrite $W_i$ as $W( \mathbf{X}_i, Z_i, \mathbf{\beta} )$. The matching estimator $\hat{\Delta}_{MW} = \hat{\mu}_1 - \hat{\mu}_0$, which converges to $\Delta_0$ asymptotically. Similar to the inverse probability weighting \cite{Lunceford2004}, this is a one-step approach that properly accounts for uncertainty with the propensity score model. The stratification and matching methods used in practice are typically two-step approaches: the propensity score model is fit in the first step, and treatment effect is estimated in a second step without adjusting for the uncertainty and correlation in estimated propensity scores in the first step.

The variance of the matching weight estimator is calculated from the sandwich method as $\widehat{\mathrm{var}}( \hat{\Delta}_{MW} ) = n^{-1}\mathbf{A}_n^{-1} \mathbf{B}_n \mathbf{A}_n^{-T}$, with $\mathbf{A}_n = n^{-1}\sum_{i=1}^n \partial \mathbf{\phi}_i(\mathbf{\theta})/\partial \mathbf{\theta}$ and $\mathbf{B}_n = n^{-1} \sum_{i=1}^n  \mathbf{\phi}_i(\mathbf{\theta}) \mathbf{\phi}_i(\mathbf{\theta})^T $. One issue remains to be resolved. Since the matching weight function (\ref{eq:MW}) does not have continuous derivative at $e_i = 0.5$, $W(\mathbf{X}_i, Z_i, \beta)$ is not everywhere differentiable with respect to $\beta$. Since (\ref{eq:MW}) equals to $\eta_1(e) = \mathrm{min}(1-e, e)/e$ when $Z = 1$ and $\eta_0(e) = \mathrm{min}(1-e, e)/(1-e)$ when $Z=0$, we solve this problem by replacing the middle piece in $\eta_1(e)$ and $\eta_0(e)$ around $0.5$ with a cubic polynomial that connects smoothly with the two ends. The result is an approximate matching weight function with continuous first derivative everywhere, which satisfies the usual regularity conditions for sandwich variance estimation. Since the middle piece can be made arbitrarily small, the approximation is quite accurate.

We first approximate $\eta_1(e)$. Let $\eta_1^*(e) = \eta_1(e)$ if $e \in (0, 0.5-\delta) \cup (0.5+\delta, 1)$ and $\eta_1^*(e) = a_0 + a_1 e + a_2 e^2 + a_3 e^3$ if $e \in [0.5-\delta, 0.5+\delta]$. In order for $\eta_1^*(e)$ to have continuous first derivative everywhere and adequately approximate $\eta_1(e)$, $\{ a_0, a_1, a_2, a_3\}$ must satisfy four conditions: (1) $\eta_1^*(0.5-\delta)=1$ (2) $\eta_1^{*'}(0.5-\delta) = 0$ (3) $\eta_1^*(0.5+\delta)=(1-2\delta)/(1+2\delta)$ (4) $\eta_1^{*'}(0.5+\delta) = -4/[(1+2\delta)^2]$. Here notation $f^{'}(.)$ denotes the first derivative of the function. Solving these four equations for $a_0$-$a_3$, we have:
$$
(a_0, a_1, a_2, a_3)^T = \mathbf{D}^{-1} \Bigl(1, 0, \frac{1-2\delta}{1+2\delta}, \frac{-4}{(1+2\delta)^2} \Bigr)^T
$$
with
$$
\mathbf{D} = \left(
               \begin{array}{cccc}
                 1 & ~~~~ 0.5-\delta & ~~~~ (0.5-\delta)^2 & ~~~~ (0.5-\delta)^3 \\
                 0 & ~~~~ 1 & ~~~~ 2(0.5-\delta) & ~~~~ 3(0.5-\delta)^2 \\
                 1 & ~~~~ 0.5+\delta & ~~~~ (0.5+\delta)^2 & ~~~~ (0.5+\delta)^3 \\
                 0 & ~~~~ 1 & ~~~~ 2(0.5+\delta) & ~~~~ 3(0.5+\delta)^2 \\
               \end{array}
             \right).
$$
Similarly, we can define $\eta_0^*(e) = \eta_0(e)$ if $e \in (0, 0.5-\delta) \cup (0.5+\delta, 1)$ and $\eta_0^*(e) = b_0 + b_1 e + b_2 e^2 + b_3 e^3$ if $e \in [0.5-\delta, 0.5+\delta]$. Then $\eta_0^*(e)$ approximates $\eta_0(e)$ with
$$
(b_0, b_1, b_2, b_3)^T = \mathbf{D}^{-1} \Bigl(\frac{1-2\delta}{1+2\delta}, \frac{4}{(1+2\delta)^2}, 1, 0 \Bigr)^T.
$$
In all the numerical studies in this paper, we set $\delta = 0.002$.

Balance diagnosis, i.e., checking whether $\mathbf{X}_i \perp Z_i | e_i $ holds, is an attractive feature of the propensity-score analysis in comparison with direct regression on the outcome \cite{Hill, Rubin1997}. In the propensity score matching paradigm, the general recommendation is to calculate the standardized difference, i.e., the absolute difference in mean divided by a pooled variance \cite{Austin2008}. If the standard difference becomes small enough after matching, that would suggest balance. However, there has been no widely accepted guidance on ``how small is being small enough''. Another issue is that although the standardized difference is very similar to the t-statistic, one can not use it to test for balance, because the sample size reduction after matching alone could reduce the significance of these tests. Exceptions to this rule exist \cite{Hansen}. In the inverse probability matching paradigm, confounders should be balanced after weighting. However, when some propensity scores are close to $0$ or $1$, the balance diagnosis statistics may be highly variable and balance is difficult to ascertain.

With the matching weights, we argue that balance can be checked with statistical tests. The presumption is that if the propensity score model is correct, then the confounders, weighted by the matching weights, should be balanced between the treated and control groups; on the other hand, if the propensity score model is misspecified, imbalance may show up in some confounders. This is like a check for propensity score model mis-specification and the null hypothesis is that the model is correctly specified. Let $X_i$ be a confounder whose balance we want to examine and $g(X_i)$ a pre-defined function of $X_i$. The balance diagnostic statistic is:
$$
\hat{B} = \frac{ \sum_{i=1}^n W_i Z_i g(X_i) }{ \sum_{i=1}^n W_i Z_i } - \frac{ \sum_{i=1}^n W_i (1-Z_i) g(X_i) }{ \sum_{i=1}^n W_i (1-Z_i) }
$$
If the propensity score model (\ref{eq:PS}) is correctly specified, each of the two terms in this expression converges to their means with respect to the maximal balanced subpopulations, denoted by $\mu_{B1}$ and $\mu_{B0}$, and they are identical. Hence, we would expect $\hat{B}$ to converge to $0$ as $n \rightarrow \infty$. To study its variance, we can again formulate the estimating equation as:
$$
0 = \sum_{i=1}^n \left( \begin{array}{c}
                                         W(\mathbf{X}_i, Z_i, \beta) Z_i ( g(X_i) - \mu_{B1}) \\
                                         W(\mathbf{X}_i, Z_i, \beta) (1-Z_i) ( g(X_i) - \mu_{B0}) \\
                                         \frac{ Z_i - e(\mathbf{X}_i, \mathbf{\beta}) }{ e(\mathbf{X}_i, \mathbf{\beta})\bigl[ 1 - e(\mathbf{X}_i, \mathbf{\beta}) \bigr] } e_{\mathbf{\beta}}(\mathbf{X}_i) \\
                                       \end{array}
                                     \right)
$$
and the variance of $\hat{B}$ follows from sandwich method. We may set $g(x) = x$ for checking any imbalance in mean, or set $g(x) = x^2$ for checking imbalance in the second moment, etc. This estimating equation can take vector-valued $g(.)$ and $X_i$ so that several confounders can be considered jointly. We can also define $g(X_{1i}, X_{2i}) = X_{1i}X_{2i}$ to study any imbalance in the correlation structure.

One caveat with theses tests is that, even when the propensity score model is correct, if many tests are performed, some tests will be significant by chance. Therefore, one should be cautious when many tests are used simultaneously. At the very least, the significance thresholds of these tests offer the data analyst a rough benchmark on whether the imbalance is small enough. If many confounders turn out to be significant, it may be a sign that the propensity score model needs improvement. It may be possible, still within the M-estimation framework, to develop an overall test of imbalance with properly controlled type I error. That is an on-going work and beyond the scope of this paper.

Nearly three decades after the propensity score was proposed, there is still ``rampant lack of good practice in propensity score matching applications'' \cite{Hill}. There are a number of reasons. First, the asymptotic theory of propensity score matching is non-standard and very complicated \cite{Abadie2006}, making it difficult to study its properties or develop theoretically justified guidelines. For example, matching depends on caliper choice but theoretically justified optimal caliper still needs to be developed \cite{AustinCaliper}. Second, matching may introduce correlation between the same matched pairs, fitting a propensity score model in the first step may introduce correlation between different matched pairs, and complicated dependence may also arise if matching is done with replacement. These correlations are often ignored or inadequately adjusted during the analysis of the matched data, and accurate variance estimation is often not available. General-purpose variance estimation methods, such as the bootstrap, does not apply to matching estimators \cite{Abadie2008}. As a result, there remains debate on whether unpaired or paired analysis of the outcome is more appropriate for matched data \cite{Austin2008, Stuart2008}.

The method of matching weights resembles the propensity score matching, but its asymptotic theory is much simpler. It is a one-step approach so that the propensity score and the treatment effect can be estimated simultaneously from a set of estimating equations, and accurate analytical variance formula is available and justified. It is somewhat subjective to choose the matching algorithm or caliper and the practice varies among data analysts. This task is no longer needed with the matching weight method.

Under homogeneous conditions, the matching weight estimator and inverse probability weighted estimator are similar when the propensity scores are not too close to $0$ or $1$. The histogram of the propensity scores concentrates more in the middle range of unit interval (Figure \ref{fig1}). When some propensity scores become extreme, the two estimators begin to diverge and the matching weight estimator is more efficient. This is observed in the simulation of Section \ref{sec:simu}. Under heterogeneous conditions, the two estimators have different estimands and are not comparable. In such case, an estimator obtained from one study may not be generalizable to other study populations. What is more likely to be generalizable, is the result of a test of the null: ``the treatment effect is zero for all subjects'' versus the alternative ``the treatment effect varies among subjects''. Assuming the statistical power is always adequate and the treatment truly affects the outcome, then the treatment effect should shown up, more or less, in different studies, regardless of the population on which the average treatment effect is defined. The test can be performed with the inverse probability weighting method and with the entire subject population as the target population. It can also be performed with the matching weight method and with the maximal balanced subpopulation as the target population. These tests will be studied in Section \ref{sec:simu}. The discussion above assumes that under the alternative the treatment effect is either positive and varies, or negative and varies among subjects. If the treatment effect is positive on some subjects and negative on other subjects, then neither inverse probability weighting, nor matching weights can guarantee enough statistical power under the alternative.

\section{Double Robust Matching Weights Estimator}
\label{sec:DR}

In this section we develop an augmented matching weights estimator that has the ``double robust'' property. The idea follows from the inverse probability weighted double robust estimator \cite{BangRobins, Lunceford2004}. In addition to the propensity score model, the augmented estimator involves two outcome models, one for the regression of $Y_i$ on $\mathbf{X}_i$ among the treated subjects, and one for the controls. Let $\mathbf{\alpha}_1$ be the parameters associated with the outcome model for the treatment group, we write $m_1( \mathbf{X}_i, \mathbf{\alpha}_1 ) = E(Y_i | \mathbf{X}_i, Z_i = 1)$ as the conditional expectation of the outcome given the covariates, and write $\mathbf{S}_1( Y_i, \mathbf{X}_i, \alpha_1)$ as the unbiased estimating equation for $\alpha_1$ derived from the likelihood or quasi-likelihood of this outcome model. For the control group, we define notation $m_0( \mathbf{X}_i, \mathbf{\alpha}_0 )$ and $\mathbf{S}_0( Y_i, \mathbf{X}_i, \alpha_0)$ similarly.

The augmented matching weights estimator is:
\begin{equation}
\begin{array}{rl}
\hat{\Delta}_{\mathrm{MW,DR}} = & ~ \dfrac{ \sum_{i=1}^n W_i \bigl\{ m_1(\mathbf{X}_i, \alpha_1) - m_0(\mathbf{X}_i, \alpha_0) \bigr\} }{ \sum_{i=1}^n W_i } ~ + ~ \\
   & ~~~~~ \dfrac{ \sum_{i=1}^n W_i Z_i \bigl\{ Y_i - m_1(\mathbf{X}_i, \alpha_1) \bigr\} }{ \sum_{i=1}^n W_i Z_i } - \dfrac{ \sum_{i=1}^n W_i (1-Z_i) \bigl\{ Y_i - m_0(\mathbf{X}_i, \alpha_0) \bigr\} }{ \sum_{i=1}^n W_i (1-Z_i) }
\end{array}  \label{eq:MWDR1}
\end{equation}

\textbf{Proposition 3.} \textit{ The augmented matching weight estimator $ \hat{\Delta}_{\mathrm{MW,DR}} $ is consistent for $\Delta_0$, as long as at least one of the following two models are correctly specified: (1) the propensity score model (\ref{eq:PS}); (2) the outcome models $m_1( X, \alpha_1)$ and $m_0( X, \alpha_0)$. } 

\textbf{Proposition 4.} \textit{ Assume that the propensity score model (\ref{eq:PS}) is known and let $\Psi = E\{ \mathrm{min}(1-e_i, e_i) \}$. The class of influence functions of regular asymptotically linear estimators for $\Delta_0$ is given by (subscript $i$ suppressed)
$$
\Bigl\{ \frac{\mathrm{min}(1-e, e)}{\Psi}\Bigl[ \frac{ZY}{e} - \frac{(1-Z)Y}{1-e} \Bigr] - \Delta_0 \Bigr\} + \Lambda
$$
where $\Lambda$ is the space of functions of form $\{ Z - e \}h(\mathbf{X})$ for any function $h(\mathbf{X})$. Among all estimators with influence functions in this class, the augmented matching weight estimator is the most efficient in the sense that it has the smallest variance. } 

The proof is similar to $\S 13.5$ of \cite{TsiatisBook}. The property described in Proposition 3 is called double robustness. In usual statistical models, if a model is misspecified, the result is usually biased. With double robustness, even if one part of the model fails, we may still get unbiased estimator with the other part of the model. Therefore, it gives the data analyst two chances, instead of one, to get a correct result. Proposition 4 shows the benefit of adding the outcome models: we will arrive at a more efficient estimator.

Double robustness has been established for inverse probability weighting method \cite{Lunceford2004}, but not for the other propensity score analysis methods. Ho et al \cite{Ho2007} and Stuart \cite{Stuart2010} mentioned that doing a regression of the outcome with propensity score matched data leads to double robust estimation, but they did not give any theoretical justification of that claim. Proposition 3 can be used to support that claim, given the similarity between the matching weight method and matching.

Estimator (\ref{eq:MWDR1}) involves unknown parameters $\alpha_1$, $\alpha_0$, and $\beta$. In practice they can be replaced by consistent estimators and Proposition 3 still holds. Let $\mu_1$, $\mu_2$, and $\mu_3$ be the asymptotic limits of the three terms in (\ref{eq:MWDR1}). The estimating equations for $(\mu_1, \mu_2, \mu_3, \alpha_1^T, \alpha_0^T, \beta^T)^T$ are:
$$
\mathbf{0} = ~ \sum_{i=1}^n \left(
\begin{array}{c}
W(\mathbf{X}_i, Z_i, \mathbf{\beta}) \{ m_1(\mathbf{X}_i, \alpha_1) - m_0(\mathbf{X}_i, \alpha_0) - \mu_1 \}  \\
W(\mathbf{X}_i, Z_i, \mathbf{\beta}) Z_i \{ Y_i - m_1(\mathbf{X}_i, \alpha_1) - \mu_2 \}  \\
W(\mathbf{X}_i, Z_i, \mathbf{\beta}) (1-Z_i) \{ Y_i - m_0(\mathbf{X}_i, \alpha_0) - \mu_3 \}  \\
\mathbf{S}_1( Y_i, \mathbf{X}_i, \alpha_1) \\
\mathbf{S}_0( Y_i, \mathbf{X}_i, \alpha_0) \\
\frac{ Z_i - e(\mathbf{X}_i, \mathbf{\beta}) }{ e(\mathbf{X}_i, \mathbf{\beta})\bigl[ 1 - e(\mathbf{X}_i, \mathbf{\beta}) \bigr] } e_{\mathbf{\beta}}(\mathbf{X}_i)
\end{array} \right)
$$
Solving these estimating equations jointly, we have $\hat{\Delta}_{MW, DR} = \hat{\mu}_1 + \hat{\mu}_2 - \hat{\mu}_3$. The variance of $\hat{\Delta}_{MW, DR}$ can be calculated by the sandwich method.

\section{Numerical Studies}
\label{sec:simu}

We conducted simulations to study the numerical performance of the matching weight estimator and double robust matching weight estimator, and compare them with three other types of propensity score analysis methods: stratification, matching, and inverse probability weighting. For stratification, we used five strata, as this is a popular choice in data analytical practice \cite{DAgostino1998, RosenbaumRubin1984}. For propensity score matching, we used the R package MatchIt (http://gking.harvard.edu/matchit) and the optimal caliper size recommended by Austin \cite{AustinCaliper}, which equals to $0.2$ times the standard deviation of the propensity score on its logit scale. To study the sensitivity of matching results to caliper size, we also consider a smaller caliper at $0.1$ and a bigger caliper at $0.3$. For inverse probability weighting (IPW), we studied the double robust IPW estimator and IPW3 \cite{Lunceford2004}. The IPW3 is the inverse probability weighting with stabilized weights, which has improved efficiency and numerical stability over simple IPW methods.

The propensity score model is a logistic regression $\mathrm{logit}\{ \mathrm{Pr}( Z_i = 1 | \mathbf{X}_i ) \} = \mathbf{X}_i^T \mathbf{\beta}$. The outcome model is a linear regression $Y_i = \Delta Z_i + \mathbf{X}_i^T \mathbf{\alpha} + \epsilon_i$ with $\Delta = 2$, $\mathbf{\alpha} = (1, 2, -1, -2, 1)^T$ and $\epsilon_i \sim N(0, 2^2)$. $\mathbf{X}_i = (X_{0i}, X_{1i}, X_{2i}, X_{3i}, X_{4i})^T$. $X_{0i} \equiv 1$. $X_{1i}$ and $X_{2i} \sim N(0,1)$. $X_{3i}$ and $X_{4i} \sim 2 \times \mathrm{Bernoulli}(0.5)$. $X_{1i}$-$X_{4i}$ are independent. We consider three scenarios: (1) $\mathbf{\beta} = (-1, 0.4, 0.2, 0.4, 0.2)^T$; (2) $\mathbf{\beta} = (-2, 0.8, 0.4, 0.8, 0.4)^T$; (3) $\mathbf{\beta} = (-3, 1.5, 0.75, 1.5, 0.75)^T$. Under these scenarios, the proportion of subjects with $Z=1$ is between $35\%$ and $40\%$, and $\mathrm{var}(Y_i)/\mathrm{var}(\epsilon_i)$ is between $3.5$ and $3.8$. The sample size is $n=1000$, and each simulation is based on $1000$ Monte Carlo replicates.

\begin{figure}
  \includegraphics[width=6in]{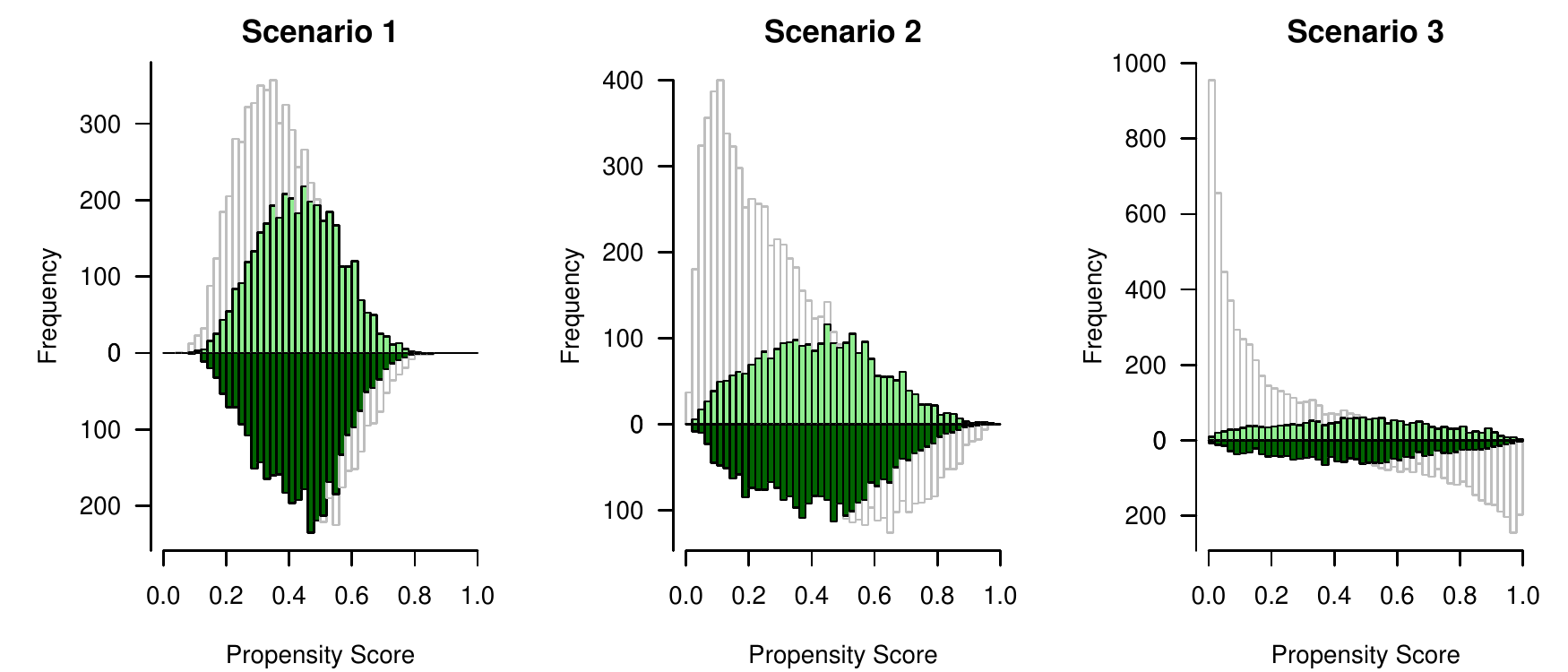}\\
  \caption{Mirror histograms illustrating the propensity scores and matching weights for the three simulation scenarios. Below horizontal zero line: $Z = 1$; above: $Z = 0$. }
  \label{fig1}
\end{figure}

The mirror histograms in Figure \ref{fig1} illustrate the difference among the three scenarios. Each mirror histogram consists of four histograms. The two outside are the ones corresponding to the distribution of propensity scores of the treated (below) and control (above) subjects. Nested within them are two histograms in color, for which each subject is weighted by $W_i$, their matching weights. Since the matching weights result in maximal balanced subgroups, the two nested histograms are like reflections in a mirror. The propensity score matching can also be illustrated in mirror histogram: the nested histograms are created by giving weight $1$ to matched subjects and $0$ to unmatched subjects. The plot is similar to Figure \ref{fig1} for the optimal caliper size and is omitted. Mirror histogram can also be used to compare continuous confounders, before and after applying the matching weights. Scenarios 1-3 represent increasing imbalance between the treated and control groups, and increasing number of subjects with propensity scores close to $0$ or $1$.

Table \ref{tab1} compares 13 point estimators. They are from: (1) true outcome regression; (2) stratification by five strata; (3) matching with caliper 0.1; (4) matching with caliper 0.2, the optimal caliper; (5) matching with caliper 0.3; (6) IPW3; (7) double robust IPW estimator; (8) matching weight (MW) estimator; (9) MW estimator with incorrect propensity score model; (10) double robust MW estimator; (11) double robust MW estimator with incorrect propensity score model; (12) double robust MW estimator with incorrect outcome model; (13) double robust MW estimator with incorrect propensity score model and outcome model. The incorrect propensity score model is the logistic regression with only $X_1$ and $X_2$ as covariates. The incorrect outcome model is the linear regression of $Y$ with $X_1$ and $X_3$ as covariates. Hence, they represent situations where some confounders are ignored.

From Table \ref{tab1} we have the following observations. First, although the matching weight method resembles matching, it is more efficient and has less bias than than the matching estimator, though its effective sample size is smaller than the sample size of the matched data set. If there are several subjects within the caliper of the propensity score, the matching algorithm chooses the one with the closest propensity score and may discard others. This is like giving weight $1$ to the matched subject and giving no weight to others. The matching weight method retains every subject in the neighborhood, and giving them roughly equal weights so that they all contribute to averaging the outcome. This increases numerical stability and efficiency, and reduces bias. Second, the results from IPW methods are comparable with the MW methods in Scenario 1, when the treated and control groups have nearly balanced propensity score distributions. This is also the situation where the propensity score can not be too close to $0$ or $1$. However, with moderate (Scenario 2) and severe (Scenario 3) imbalance, the IPW methods may suffer from large loss of efficiency. In fact, the matching weight estimator without augmentation is even more efficient than the double robust IPW estimator in these scenarios. Third, the simulation results support the double robust and efficiency properties stated in Propositions 3 and 4. Interestingly, even when both the propensity score model and the outcome models are incorrect, the double robust matching weight estimator is still better than the matching weight estimator with incorrect propensity score model and without augmentation. Therefore, it seems that the double robust matching weight estimator should always be recommended in data analytical practice. Finally, the stratification method clearly has larger bias than other methods, and is not efficient in general. This observation agrees with that in \cite{Lunceford2004}. The stratification does not produce consistent estimators, and there is still a lack of guidance on whether the number of strata should increase with the sample size. Table \ref{tab2} shows the empirical coverage probabilities of $95\%$ confidence interval of various matching weight estimators. The result indicates that the sandwich variance formula is accurate.

We conducted a simulation under the heterogeneous treatment effects, based on the discussion at the end of $\S 2$. We modify the outcome model as $Y_i = \Delta_i Z_i + \mathbf{X}_i^T \mathbf{\alpha} + \epsilon_i$, with $\Delta_i = \theta( 2.5 + 0.5 X_{1i} - 0.5 X_{3i} )$. This setup ensures that the treatment has a positive effect on almost every subject (except $< 0.1\%$ of the cases), but the magnitude of the effect varies. The parameter $\theta$ controls the overall size of the effect. When $\theta = 0$, the treatment has no effect on the outcome. Table \ref{tab3} presents two-sided $0.05$-level test of the null hypothesis $\theta = 0$. The test statistic is a Wald type statistic with the matching weight estimator divided by its sandwich standard error. The type I error is correct even when the sample size is as low as 200. In comparison, the double robust IPW estimator has inflated type I error and lower statistical power at these sample sizes. The rejection probability under the alternative likely depends on how the treatment effect varies with subjects, and hence it is difficult to study theoretically and draw generalizable conclusions.

\bibliographystyle{plain}
\bibliography{MWpaper-ref}

\newpage
 
\begin{table}
\begin{center}
\begin{tabular}{lcccccccccccc} \\
  \hline
       &  \multicolumn{4}{c}{Scenario 1} &  \multicolumn{4}{c}{Scenario 2}  & \multicolumn{4}{c}{Scenario 3} \\
  Method      & bias  & var & MSE  & ESS & bias  & var & MSE  & ESS & bias  & var & MSE   & ESS   \\
  \hline
  1:best      & 0.1   & 100 & 100  & -   & 0.7   & 100 & 100  & -   & 0.1   & 100 & 100   & - \\
  2:strt      & 2.6   & 115 & 131  & -   & 4.9   & 185 & 227  & -   & 6.4   & 546 & 612   & - \\
  3:M 0.1     & 0.7   & 183 & 184  & 726 & 1.1   & 217 & 217  & 538 & -1.3  & 285 & 287   & 390 \\
  4:M opt     & 0.6   & 170 & 170  & 758 & 1.3   & 191 & 193  & 571 & -3.2  & 244 & 261   & 429 \\
  5:M 0.3     & 0.4   & 164 & 164  & 784 & 1.5   & 197 & 199  & 603 & -5.7  & 231 & 284   & 469 \\
  6:IPW3      & 0.1   & 110 & 110  & -   & 1.0   & 199 & 200  & -   & 4.0   & 512 & 538   & - \\
  7:DR IPW    & 0.0   & 103 & 103  & -   & 0.4   & 146 & 145  & -   & 0.1   & 494 & 494   & - \\
  8:MW        & 0.1   & 106 & 106  & 714 & 0.6   & 115 & 114  & 520 & 0.1   & 130 & 130   & 366 \\
  9:MW p      & -29.8 & 225 & 2254 & -   & -55.0 & 197 & 5723 & -   & -87.0 & 157 & 12652 & - \\
  10:DR MW    & 0.1   & 102 & 102  & -   & 0.6   & 105 & 105  & -   & 0.2   & 113 & 113   & - \\
  11:DR MW p  & 0.0   & 101 & 101  & -   & 0.6   & 106 & 106  & -   & 0.3   & 106 & 106   & - \\
  12:DR MW y  & 0.1   & 104 & 104  & -   & 0.6   & 108 & 108  & -   & 0.2   & 116 & 116   & - \\
  13:DR MW py & 9.4   & 126 & 327  & -   & 17.9  & 127 & 708  & -   & 25.7  & 130 & 1217  & - \\
  \hline
\end{tabular} 
\end{center}
\caption{Compare estimators on bias, variance, mean squared error (MSE) and effective sample size (ESS). Bias is expressed as \% difference from the true value $\Delta=2$. Variance and MSE are expressed as a percentage of those of Method 1.}
\label{tab1}
\end{table}

\begin{table}[t]
\begin{center}
\begin{tabular}{lcccc} \\
  \hline
                 &  \multicolumn{3}{c}{Scenario}  \\
  Method         &  1   &  2  &   3  \\
  \hline
  8:MW          & 94.2 & 93.9 & 95.0 \\
  9:MW p      & 16.0 & 0.1 & 0.0 \\
  10:DR MW      & 94.1 & 94.4 & 94.8 \\
  11:DR MW p  & 94.3 & 94.7 & 95.1 \\
  12:DR MW y   & 94.2 & 94.3 & 94.4 \\
  13:DR MW py & 74.3 & 48.2 & 24.8  \\
  \hline
\end{tabular} 
\end{center}
\caption{Coverage probabilities (\%) of matching weight estimator}
\label{tab2}
\end{table}

\begin{table}[b]
\begin{center}
\begin{tabular}{lcccccc}
\hline
          &  \multicolumn{2}{c}{8:MW} &  \multicolumn{2}{c}{10:DR MW}  & \multicolumn{2}{c}{7:DR IPW} \\
$\theta$  &  $n=200$  &  $n=600$ &  $n=200$  &  $n=600$ &  $n=200$  &  $n=600$  \\ 
\hline
0  & 4.7 & 4.7 & 5.2 & 5.1 & 6.8  &  7.5 \\
0.25 & 24.6 & 65.4 & 30.9 & 67.3 & 28.2 & 61.4 \\
0.50 & 75.2 & 99.9 & 79.5 & 99.8 & 75.3 & 98.4 \\
\hline
\end{tabular} 
\end{center}
\caption{Rejection probabilities (\%) under heterogeneous conditions (Scenario 2)}
\label{tab3}
\end{table}

\end{document}